\documentclass[twocolumn,tighten,usenames,dvipsnames,twocolappendix]{aastex631}

\usepackage{amsmath}
\usepackage{amsfonts}
\usepackage{eht}

\shorttitle{Learning to See}
\shortauthors{Kouroshnia et al.}

\begin{document}
\submitjournal{The Astrophysical Journal}
\title{Learning to See: Applying Inverse Recurrent Inference Machines to See through Refractive Scattering}

\author[0009-0002-6575-2492]{Arvin Kouroshnia}
\affiliation{Waterloo Centre for Astrophysics, University of Waterloo, Waterloo, ON, N2L 3G1, Canada}
\affiliation{Waterloo Centre for Astrophysics, University of Waterloo, Waterloo, ON, N2L 3G1, Canada}
\affiliation{Department of Physics and Astronomy, University of Waterloo, 200 University Avenue West, Waterloo, ON, N2L 3G1, Canada}

\author[0009-0004-4507-5230]{Kenny Nguyen}
\affiliation{Waterloo Centre for Astrophysics, University of Waterloo, Waterloo, ON, N2L 3G1, Canada}
\affiliation{Department of Physics and Astronomy, University of Waterloo, 200 University Avenue West, Waterloo, ON, N2L 3G1, Canada}

\author[0000-0003-1361-5699]{Chunchong Ni}
\affiliation{Perimeter Institute of Theoretical Physics, 31 Caroline Street North, Waterloo, ON, N2L 2Y5, Canada}
\affiliation{Department of Physics and Astronomy, University of Waterloo, 200 University Avenue West, Waterloo, ON, N2L 3G1, Canada}
\affiliation{Waterloo Centre for Astrophysics, University of Waterloo, Waterloo, ON, N2L 3G1, Canada}

\author[0009-0003-4620-8448]{Ali SaraerToosi}
\affiliation{Perimeter Institute of Theoretical Physics, 31 Caroline Street North, Waterloo, ON, N2L 2Y5, Canada}
\affiliation{Department of Physics and Astronomy, University of Waterloo, 200 University Avenue West, Waterloo, ON, N2L 3G1, Canada}
\affiliation{Waterloo Centre for Astrophysics, University of Waterloo, Waterloo, ON, N2L 3G1, Canada}
\affiliation{Department of Computer Science, University of Toronto, 40 St. George St., Toronto, ON, M5S 2E4, Canada}

\author[0000-0002-3351-760X]{Avery E. Broderick}
\affiliation{Perimeter Institute of Theoretical Physics, 31 Caroline Street North, Waterloo, ON, N2L 2Y5, Canada}
\affiliation{Department of Physics and Astronomy, University of Waterloo, 200 University Avenue West, Waterloo, ON, N2L 3G1, Canada}
\affiliation{Waterloo Centre for Astrophysics, University of Waterloo, Waterloo, ON, N2L 3G1, Canada}

\begin{abstract}
The Event Horizon Telescope (EHT) has produced horizon-resolving images of Sagittarius A* (\sgra).  Scattering in the turbulent plasma of the interstellar medium distorts the appearance of \sgra on scales only marginally smaller than the fiducial resolution of EHT. Therefore, this process both diffractive blurs and adds stochastic refractive substructures that limits the practical angular resolution of EHT images of \sgra. We utilized a novel recurrent neural network machine learning framework to demonstrate that it is possible to mitigate interstellar scattering at wavelengths of $1.3\,{\rm mm}$ near the galactic center up to structures at the scale of $5\mu{\rm as}$ well below the nominal instrumental resolution of EHT, $24\,\mu{\rm as}$.
\end{abstract}


\section{Introduction}
\label{Introduction}

Sagittarius A* (\sgra) has now been imaged on angular scales sufficient to resolve the event horizon by the Event Horizon Telescope \citep[EHT;][hereafter \citetalias{PaperI,PaperIII}]{M87PaperII, PaperI, PaperIII}.  The images of \sgra have yielded a wealth of information about the black hole and surrounding accretion flow, including its mass and the detection of magnetohydrodynamic turbulence \citep[][hereafter \citetalias{PaperIV,PaperV}]{PaperIV,PaperV}.  These conclusions rest upon the accurate identification of high-resolution features in the EHT images.  
At a wavelength of $\lambda=1.3\,{\rm mm}$, a global very long baseline interferometry (VLBI) experiment, like EHT, has a nominal resolution of $24\,\mu{\rm as}$ and an effective super-resolution of roughly factor of two better \citepalias[see ][]{PaperIII}.  Shorter wavelengths result in higher resolutions; at $\lambda=0.87\,{\rm mm}$ the EHT nominal resolution is $14\,\mu{\rm as}$.  

The imaging of Sgr A* by EHT faced several significant challenges, including the limited number of baselines, intrahour intrinsic variability, and interstellar scattering \citepalias{PaperIII,PaperIV}.  The last, scattering, is a well-known effect that impacts all radio sources near the Galactic center \citep{1976MNRAS.177..319D}.  Scattering both blurs images and introduces additional substructures on scales marginally smaller than the resolution of EHT at $1.3\,{\rm mm}$, and represents a fundamental limit for EHT and future high-resolution imaging experiments.  The origin of the scattering is believed to be stochastic fluctuations in the density free electrons present in the interstellar medium \citep[ISM;][]{Bower_2014}.

The magnitude of the impact of interstellar scattering depends on wavelength: diffractive blurring $\propto\lambda^2$ and $\propto\lambda^4$, and therefore shorter wavelengths are less impacted.  However, atmospheric absorption places practical limits ground-based high-frequency VLBI observations to $\lambda\ge0.87\,{\rm mm}$; for shorter wavelengths the number of contemporaneously available stations is small even with new techniques \citep[e.g.,][]{fpt}.  Therefore, VLBI observations at $\lambda\le1.3\,{\rm mm}$ will remain a staple of horizon-resolving observations of \sgra well into the future.  Hence, scattering presents a recurring difficulty for the interpretation of EHT images.

Interstellar scattering produces two types of distortion at millimeter wavelengths: It diffractively blurs the image and adds refractive noise. The former, in principle, is entirely reversible, while the latter is not. The EHT handles these two effects separately. It deblurs the image using the method outlined in \citet{Fish_2014, Bower_2014, Broderick_2009} and it adds systematic uncertainties to handle the noise \citepalias{PaperIII,PaperIV}.  Although refractive noise removes information, it also has specific properties that can aid in its mitigation.  It has a known power spectrum \citep{Johnson_2018, Gwinn_2014}, presumably arising from the turbulent spectrum of ISM density fluctuations.  In addition, it is non-birefringent, i.e., it impacts all polarization modes in a similar fashion \citep{Ni_2022}. 

In this paper, we demonstrate that interstellar scatting can be effectively mitigated at all resolutions relevant for ground-based high-frequency VLBI experiments, like EHT.  We do this by using a machine learning algorithm to deblur and remove refractive substructures, ultimately allowing most features of the intrinsic structure of Sgr A* to be recovered. Our model leverages the known physics of the scattering screen (i.e., the difference in the power spectra between refractive and intrinsic substructures, and the lack of birefrigence in the ISM at mm wavelengths).  Importantly, it does not make strong assumptions regarding the global structure of the intrinsic images (e.g., we do not adopt a ring prior on the intrinsic image). 

We utilizes the Invertible Recurrent Inference Machine (IRIM) architecture \citep{pputzky2019}, which is an invertible variant of Recurrent Inference Machines (RIM) developed by \citep{putzky2017recurrentinferencemachinessolving} to denoise images.  The IRIM model has constant memory complexity, and it is a U-Net architecture which applies dilated convolution kernels which will detect structures at different scales (see, e.g., \cite{ronneberger2015unetconvolutionalnetworksbiomedical}).  The U-Net discussed in RIM is an extension of the models discussed in \citet{Zheng_2015}, \citet{46f49a297c8542aa80d7fad85aed17e0}, \citet{NIPS2016_f4be0027}, and \citet{chen2015learningoptimizedreactiondiffusion}, which is useful in training on non-convex optimization problems. We train the IRIM model on Gaussians with superimposed fluctuations (Kolmogorov Gaussians), and reserve general relativistic magnetohydrodynamic (GRMHD) simulations for test images \citep{Porth_2019}.

In \autoref{Background}, we describe the physics of scattering and then propose the use of IRIM to mitigate it. In \autoref{Method}, we elaborate on the generation of training data and model training. In \autoref{Discussion}, we reportthe model's performance in mitigating scattering for the Kolmogorov Gaussians and test this model on GRMHD simulations.  Conclusions are collected in \autoref{Conclusion}.

\section{Background}
\label{Background}

\subsection{Interstellar Scattering}
\label{Insterstellar Scattering}

\begin{figure}
    \centering
    \includegraphics[width=0.99\columnwidth]{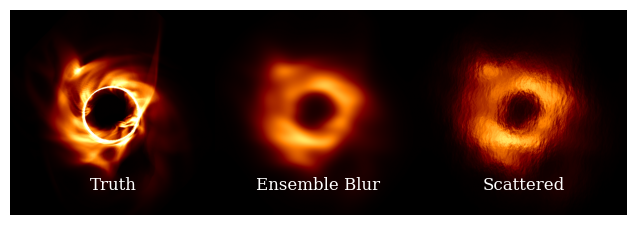}
    \caption{An example of the effects of interstellar scattering.
    For comparison, the unscattered image $I(\textbf{r})$ (left), the image after convolution with the diffractive scattering kernel $\langle I \rangle(\textbf{r})$ (center), and the fully scattered image $\mathcal I(\textbf{r})$ (right) are shown for a field of view of  $200\,\mu{\rm as}$.  In practice, only the fully scattered image may be observed.  Note that the brightness scale for each image is chosen independently.}    
    
    \label{fig:scattering-label}
\end{figure}

The inhomogeneity of the ISM, especially the free electron clouds, leads to extra phase in the EM wave propagating from the galactic centre, causing scattering. This process has two effects on the VLBI data \citep{rickett1984slow}. First, scattering produces a diffractive blur \citep{Issaoun_2021}. This is shown in the center image of \autoref{fig:scattering-label}. The blur broadens proportional to $\lambda^2$, where $\lambda$ is the observing wavelength in centimeters \citep{Bower_2006}. This blur is anisotropic; along one direction the scatter-broadening kernel is nearly twice as large as the other \citep{Cho_2022}. The full-width half mass (FWHM) of the kernel is
\begin{equation}
\begin{split}
 \rm FWHM_{maj}  &= (1.380 \pm 0.013) (\lambda/1\,{\rm cm})^2 \, {\rm mas}
\\  \rm FWHM_{min}  &= (0.704 \pm 0.013) (\lambda/1\,{\rm cm})^2 \,{\rm mas}.
\end{split}
\label{eq:scatt_fwhm}
\end{equation}
Interstellar scattering also adds refractive noise which adds additional substructures. This is shown in the right image of \autoref{fig:scattering-label}.

Interstellar scattering can be well approximated using a thin screen \citep{goodman1989shape, narayan1989shape}. The radio waves propagate through space. It then interacts with a screen between the source and the observer. This changes the phase of the wave by an amount $\phi(\textbf{r})$ where $\textbf{r}$ is the transverse position. To analyze the statistical properties of $\phi(\textbf{r})$, we define the \textit{phase structure function} as $D_\phi \equiv \langle [\phi(\textbf{r}_0 + \textbf{r}) - \phi(\textbf{r}_0)]^2 \rangle_{\textbf{r}_0}$. $\langle \dots \rangle_{\textbf{r}_0}$ represents the average over different realizations of $r_0$. The phase structure function is assumed to follow a power law, i.e. $D_{\phi}(\textbf{r}) \propto |\textbf{r}|^\alpha$ \citep{armstrong1995electron}, where $\alpha$ is assumed to be the Kolmogorov index, $5/3$ \citep{Johnson_2015}. However, observations of Sgr A* has shown that it is closer to $1.38$ \citep{Johnson_2018}. The Kolmogorov index is used in the model given that is a common power law for various kinds of noise. 

The scattered image $\mathcal{I}$ is related to the intrinsic image $I$ as follows:
\begin{equation}
\begin{split}
\mathcal{I}(\textbf{r}) &= \langle I \rangle(\textbf{r} + r_F^2 \nabla \phi(\textbf{r})) 
\\&= (K \ast I)(\textbf{r} + r_F^2 \nabla \phi(\textbf{r})).
\label{scattering-equation}
\end{split}
\end{equation}
$\langle I \rangle(\textbf{r})$ is the ensemble average image, the scattered image averaged over multiple realizations of $\phi(\textbf{r})$. Under this averaging regime, refractive noise is mitigated. The fresnel radius, $r_F = \sqrt{\frac{DR}{D+R}\frac{\lambda}{2\pi}}$ is the scale at the observer where the spherical curvature of the incoming radio wave becomes prominent. The distances D and R are the distances from the scattering screen to the observer and Sgr A*.

The effects of diffractive blurring is approximated by the ensemble average blur \citep{Johnson_2016}. Hence, diffractive scattering is entirely reversible since it does not depend on the realization of $\phi(\textbf{r})$. The ensemble average kernel $K$ is 
\begin{equation}
\begin{split}
\tilde K(\textbf{b}) = \exp\left[-\frac{1}{2}D_\phi\left(\frac{\textbf{b}}{1+M}\right)\right].  
\end{split}
\end{equation}
$\tilde K(\textbf{b})$ is the Fourier transform of $K(\textbf{x})$ and $\textbf{b}$ is the baseline position in the Fourier domain. $M \equiv D/R$, $D$ is the distance between the screen and the observer while $R$ is between the screen and the source.

The kernel $K$ is only dependent on the phase structure function $D_\phi(\textbf{r})$, not the unknown phase function $\phi(\textbf{r})$. The diffractive kernel is entirely known. At short baselines, e.g. at large scales or long observing wavelengths, $D_\phi(\textbf{r})$ is relatively small and can be assumed to be symmetric. The resulting kernel is then approximately a Gaussian.

\autoref{scattering-equation} can be linearly approximated as, 
\begin{equation}
\begin{split}
\mathcal{I}(\textbf{r}) &\approx \langle I\rangle(\textbf{r}) + r_F^2  (\nabla \langle I \rangle \cdot  \nabla \phi)(\textbf{r})
\\ &\approx (K \ast I)(\textbf{r}) + r_F^2  (\nabla (K \ast I) \cdot  \nabla \phi)(\textbf{r}).
\label{linear-approx}
\end{split}
\end{equation}
This separates the diffractive blur in the first term from the refractive noise in the second term.

Interstellar scattering is non-birefringent, i.e. $\phi(\textbf{x})$ is the same across all four polarizations, I, Q, U and V \citep{Ni_2022}.  The root mean square phase difference between two polarizations of images of objects at the Galactic Centre, e.g. Sgr A*, is 
\begin{equation}
\begin{split}
\sqrt{\langle\delta\phi^2\rangle} \approx 10^{-12} \left(\frac{B}{1 \mu G}\right) \left(\frac{r_{in}}{800 ~\rm km}\right)~{\rm rad}.
\label{non-birefrigent}
\end{split}
\end{equation}
$r_{in}$ is the inner scale which is measured to be $\approx 800\,{\rm km}$ \citep{Johnson_2018}. This is assuming that the observing wavelength is $1.3\,{\rm mm}$, the wavelength at which images of Sgr A* and M87* have been produced \citep{Event_Horizon_Telescope_Collaboration_2022_I,Akiyama_2019}.

\autoref{non-birefrigent} produces a small phase difference between the polarizations, which means that the scattering screen $\phi(\textbf{r})$ is the same between the four polarizations.

\autoref{scattering-equation} is a linear map between $I$ and $I'$ with the addition of a random component $\phi$. Hence, recovering the intrinsic $I$ from the scattered image $I'$ is an example of a noisy inverse problem. In principle, it should be possible to leverage linearity and the assumptions about scattering mentioned earlier in this section, e.g. non-birefringence and the well-studied phase structure function, to develop a strategy to mitigate scattering effectively.

\subsection{IRIM}

IRIM is a general inference tool to address the problem of recovering the intrinsic underlying data $\textbf{x}$ as accurately as possible from an observed noisy value $\textbf{y}$ that differs from $\textbf{x}$ by a known or assumed linear mapping $A$ and a noise $\epsilon$ that is generated by a known stochastic process, 
\begin{equation}
\begin{split}
y = A\textbf{x} + \epsilon.
\label{Inverse}
\end{split}
\end{equation}
Due to the noise $\epsilon$, finding a $\textbf{A}^{-1}$ is a non-trivial problem. IRIM recovers $\textbf{x}$ by iteratively refining an initial estimate $\hat{\textbf{x}}$ by comparing it to the data $\textbf{y}$.

This approach is an example of finding the maximum a posteriori,

\begin{equation}
\max_{\hat{\textbf{x}}} \left[\log p(\textbf{y}|\hat{\textbf{x}}) + \log p (\hat{\textbf{x}})\right],
\label{MAP}
\end{equation}
where $p(\textbf{y}|\hat{\textbf{x}})$ is the likelihood and $p(\hat{\textbf{x}})$ is the prior. These two functions must be defined to find the $\hat{\textbf{x}}$ that maximizes that the a posteriori in addition to a numerical method to maximize \autoref{MAP}. 

In contrast, IRIM maximizes the likelihood by having the user provide a predefined gradient function $\partial_{\textbf{y}|\textbf{x}} = \partial_{\textbf{x}} \log p(\textbf{y} \vert \textbf{x})$, where $\partial{\textbf{x}}$ is the gradient with respect to the parameters of $\textbf{x}$. The priors might not be obvious, or they might not be trivial to maximize, so IRIM uses a machine learning model to maximize them \citep{putzky2017recurrentinferencemachinessolving}. IRIM implements \autoref{MAP} as follows, 

\begin{equation}
\begin{split}
\hat{\textbf{x}}_{n+1} = \hat{\textbf{x}}_n + h_\Omega(\partial_{\textbf{y} \vert \textbf{x}} (\hat{\textbf{x}}_n), \hat{\textbf{x}}_n),
\end{split}
\end{equation}

where $n$ is the inference step, the number of times the IRIM model is applied to $\hat {\textbf{x}}$, $h$ is the IRIM model, and $\Omega$ are the machine learning parameters of $h$.

Depending on the problem, the values of $\textbf{x}$ could be constrained, e.g., ensuring an image is real or ensuring an image is positive definite.  To handle this, nonlinear link functions are used to ensure 
$\textbf{x}$ has appropriate values;
\begin{equation}
\hat{\textbf{x}} = \Psi (\eta),
\end{equation}
where $\Psi (\eta)$ is the link function and $\eta$ is the unconstrained space in which the IRIM model iterates. To enhance the inference process, a latent memory variable $s$ is introduced. This allows the IRIM to have memory, which helps in tracking progression, curvature, and other aspects of the iterative process. The update equations that incorporate the memory state are
\begin{equation}
\begin{split}
\eta_{n+1} &= \eta_n + h_\Omega (\eta_n, \ s_{n+1}) 
\\ s_{n+1} &= h_\Omega^* (\eta_n, s_n),
\end{split}
\end{equation}
where $h_\Omega$  and $h_\Omega^*$ are the update functions for the variable and the state, respectively. During training, IRIM performs N inference steps. The estimate $\hat{\textbf{x}}_N$ is then compared with the true $\textbf{x}$ during training as a loss function. The set of $\Omega$ that minimizes this loss is then calculated.

\subsection{Scattering Mitigation as an Inverse Problem}

Scattering can be treated approximately as an inverse problem when compared to \autoref{linear-approx} when compared to \autoref{Inverse}, 
\begin{equation}
\begin{split}
\textbf{y} &= \mathcal{I}(\textbf{r})
\\ \textbf{x} &= I(\textbf{r})
\\ \textbf{f}(\textbf{x}) &= (K \ast I)(\textbf{r})
\\ \epsilon &= r_F^2 [\nabla \phi \cdot \nabla (K \ast I)](\textbf{r}).
\end{split}
\end{equation}

The blurring kernel functions as the corrupting process, and the refractive contribution, the second term in \autoref{linear-approx}, functions as the stochastic process. Even though the last line of \autoref{linear-approx} is an approximation of \autoref{scattering-equation}, this should not be an issue as it has been empirically demonstrated that IRIM is able to solve the more general inverse problem,
\begin{equation}
\begin{split}
\textbf{y} = A(\textbf{x}, \epsilon),
\end{split}
\end{equation}
where $A$ is not necessarily linear and $\epsilon$ is not additive \citep{pputzky2019}. IRIM should be able to learn how to mitigate \autoref{scattering-equation}.

For this problem, a link function would not be necessary hence $\hat{S}^\mu_n = \eta_n^{\mu}$, where $\hat{S}^\mu$ are the Stokes parameters, $(I, Q, U, V)$. The initial estimate is the scattered image $\eta_0^{\mu} = \mathcal{S}^{\mu}$. In this paper, a superscript represents a value related to a specific polarization, e.g. $S^i$ represents one of the polarization maps Q, U or V. Latin superscripts exclude the Stokes parameter $I$. In contrast, Greek ones include it, i.e. $S^\mu$ could be realized as I, Q, U or V. We define the likelihood gradient function as:

\begin{equation}
\begin{split}
\label{likelihood}
\partial^\mu_{\textbf{y}|\textbf{x}} = \mathcal{S}^\mu - \hat{S}^\mu.
\end{split}
\end{equation}

\autoref{likelihood} is to ensure that $\hat{S}^\mu$ would preserve most of the structure of $\mathcal{S}^\mu$. This measures the accuracy of the images.

To use IRIM to mitigate scattering, the size of the latent memory $s_n$, the kernel size, and the dilations of the kernel will have to be defined. We define these parameters for our specific model in \autoref{Method}.

\section{Model Specifications}
\label{Method}

To train the IRIM model, we provide it with a set of ``observed'' images that have undergone the process outlined by \autoref{scattering-equation}, $\mathcal S^\mu$, and then compare it with the corresponding ``truth'' images, the $S^\mu$. The truth images are what these hypothetical structures should look like while the observed images are what the EHT would theoretically observe if it does not factor scattering into consideration and has a complete (u,v)-coverage.

`   \subsection{Architecture}

\begin{figure}
    \centering
    \includegraphics[width=0.999\linewidth]{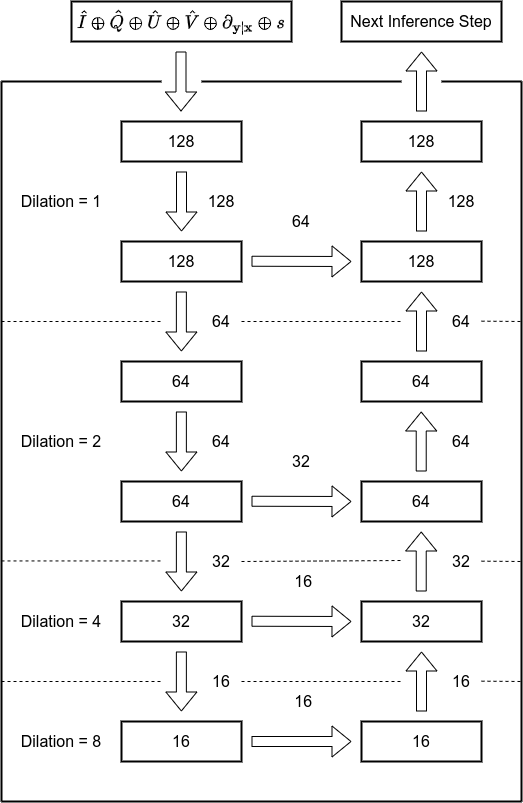}
    \caption{The image channels and dilation of the IRIM model. Each box in this diagram represents an invertible inference layer as outlined in \citet{pputzky2019}. The numbers associated with each box represent the number of image channels being processed. The arrows represent the flow of image channels to the next layer. The dilation is the spacing between the values of each kernel of each inference layer.}
    \label{model-diagram}
\end{figure}

To build our inference machine, we utilize the \texttt{IRIM}\footnote{
\url{https://github.com/pputzky/invertible_rim}} package. We modified the example denoising script as shown in \autoref{model-diagram}. In particular, we added extra layers with dilation rates of 1, 2, and 8. The increased layers of dilation rates of 1 and 2 allow the model to make more complex computations while the ones with a dilation rate of 8 allows it to recognize larger-scaled features. We also increased the number of image channels being processed from 16 to 128 to allow for more complex computation. 

 \subsection{Training Set Generation}
\label{sec:training_set}

The truth images are created using a Kolmogorov profile of a given FWHM,
\begin{equation}
\begin{split}
g(\textbf{r} )= \frac{1}{\sigma\sqrt{2\pi}}\exp\left(-\frac{|\textbf{r} - \textbf{r}_0|^2}{2\sigma^2}\right)
\label{g-envelope},
\end{split}
\end{equation}
where $\sigma = \text{FWHM}/2\sqrt{2\ln 2}$ and $\textbf{r}_0$ is the center of the profile, and a noise that is generated from a specific power spectrum. The power spectrum is the Fourier transform modulus squared of the image, i.e. $P^\mu(\textbf{b}) \equiv | \tilde S^\mu (\textbf{b})|^2$. We apply a power law to the power spectrum to generate the noise, i.e. $\langle P \rangle(\textbf{b}) \propto |\textbf{b}|^{-(\alpha+2)}$. We set the power-law index to $\alpha = 5/3$, i.e. the Kolmogorov scale, since it is a common scale that appears when analyzing turbulence, and it is chosen independent of any knowledge of GRMHD. We wanted to test if the model could generalize to images of different power-law indices.

The noise in each Stokes channel is, then, set by
\begin{equation}
\begin{split}
\tilde n^\mu(\textbf{b}) \propto \beta^\mu \exp(i\theta^\mu) \sqrt{\langle P \rangle(\textbf{b})}.
\end{split}
\end{equation}

The complex phase $\theta^\mu$ and amplitude $\beta^\mu$ are random values independently sampled from uniform and unit-variance normal distributions for each realization of $\textbf{b}$.
This noise is superimposed on the Kolmogorov profile $g(\textbf{r})$ via
\begin{equation}
\begin{split}
\label{pos-def-apply}
I \propto g\exp(n^0) \approx g|n^0+1|,
\end{split}
\end{equation}
which ensures that $I$ remains non-negative.  The constant of proportionality is chosen such that the maximum intensity is unity, i.e., $I \in [0,1]$.

The other Stokes maps
are generated from $I$ with additional noise map realizations. Because $Q$, $U$, and $V$ may be negative, we apply this polarization-specific noise via
\begin{equation}
\begin{split}
\label{non-pos-apply}
S^{i} \propto In^i,
\end{split}
\end{equation}

where $S^i$ is normalized to ensure that $S^i \in (-1, +1)$. The intensity is said to be greater than 1 as this gives a larger range for intensity such that observational intensity $I<1$ is a subset of this generalized intensity developed for this training set in which $I^2-Q^2-U^2-V^2>0$. 

We utilize the \texttt{stochasitic optics} module of the \texttt{eht-imaging}\footnote{
\url{https://github.com/achael/eht-imaging/}} package to simulate scattering. \texttt{stochastic optics} implements both the exact \autoref{scattering-equation} and the linear approximation \autoref{linear-approx}. To generate $\mathcal{S}^\mu$ from $S^\mu$, we used \autoref{scattering-equation} and set the observing wavelength to $\lambda = 1.3\,\mm$.

\subsection{Training}
We generated a set of 200,000 truth-scattered image pairs of Kolmogorov Gaussians and trained the model over one interaction of this set. We apply the IRIM model over 20 inference steps, i.e. $N = 20$ with a learning rate of $3\times 10^{-5}$. We then compare the image estimate of the last inference step, $\hat{S}^\mu_N$, with the true $S^\mu$ using mean square error (MSE), i.e.
\begin{equation}
\begin{split}
{\rm MSE} \equiv \sum_{\mu \in \{I,Q,U,V\}} \iint_A d^2r \left(\hat S^\mu_{N}(\textbf{r}) - S^\mu(\textbf{r})\right)^2.
\end{split} \end{equation}
$A$ is the area in the sky that the image occupies. We then used gradient descent to minimize this value to train the IRIM model. This is used to measure the accuracy of the machine learning parameters. A single NVIDIA GeForce RTX 2080 Ti was employed to train this model, with the training process completed in one day.

\section{IRIM Scattering Mitigation Performance}
\label{Discussion}


Here, we assess the performance of the IRIM model in mitigating scattering for multiple source images.  We do this by estimating the effective resolution to which the impacts from scattering may be effectively mitigated in a fashion similar to how image reconstruction fidelity is estimated in \citet{Akiyama_2019nxcorr}.  The effects of scattering are dependent on the underlying flux distribution, and therefore so is the performance of any scattering mitigation scheme.  Thus, we assess the ability to descatter both images similar to those in the training set and a set of general relativistic magnetohydrodynamic simulations (GRMHD) which qualitatively differ from the training set and provide a direct proxy for what might be seen by EHT and future experiments.

\subsection{Effective Resolution of Scattering Mitigation}

We expect any scattering mitigation scheme to fail at sufficiently small scales due to the suppression of small-scale structure by diffractive scattering.  Therefore, we characterize the performance of scattering mitigation with an effective resolution, determined by comparing the truth and estimated Stokes maps after convolution with Gaussians of various sizes.  This procedure is similar to that described to assess reconstruction fidelity in \citet{Akiyama_2019nxcorr}.

Specifically, to compare the truth, $S^\mu$, and estimated, $\hat{S}^\mu$, Stokes map we employ the normalized cross-correlation (NXCORR).  As in \citetalias{PaperIV}, we define the NXCORR for a given Stokes map via, $\rho^\mu$, by
\begin{equation}
\begin{split}
\rho^\mu \equiv \frac{1}{|A|}\iint_{A} d^2r \frac{(\hat S^\mu(\textbf{r}) - \langle \hat S^\mu\rangle_{\textbf{r}})(S^\mu(\textbf{r}) - \langle S^\mu\rangle_{\textbf{r}})}{\sigma_{\hat S^\mu}  \sigma_{S^\mu} },
\end{split}
\end{equation}
where $\langle \dots \rangle_{\textbf{r}}$ is the mean-value over the area of the image $A$ and $\sigma_{S^\mu}$ and $\sigma_{\hat{S}^\mu}$ are the standard deviation of $S^\mu$ and $\hat{S}^\mu$ across $A$ respectively. $\rho^\mu$ is always $\in [0,1]$ for Stokes $I$ and $\in[-1,1]$ for Stokes Q, U, and V. For all Stokes maps, $\rho^\mu=1$ only if $S^\mu$ and $\hat S^\mu$ are identical to each other, while $\rho^\mu<1$ implies differences that become large as $\rho^\mu \gtrsim 0$. While the values of $\rho^\mu$ are similar across Stokes maps for all cases that we considered, we present a combined statistic, $\rho$, given by the average over the four Stokes maps, i.e. $\rho = (1/4) \sum_{\mu \in \{I,Q,U,V\}} \rho^\mu$. 
Prior to constructing $\rho$, we apply a Gaussian blur with various FWHM to both $S^\mu$ and $\hat{S}^\mu$, effectively suppressing the impact of differences on angular scales smaller than the FWHM.  Hence, given a threshold $\rho$ value, here taken to be $\rho=0.95$, this procedure provides a direct estimate of the scales to which high-fidelity scattering mitigation may be performed.

For comparison, we also perform a similar characterization of the original scattered image, $\mathcal{S}^\mu$, and a deblurred image that removes only the effects of diffractive scattering.  This deblurred image is generated using the blurring kernel, $K$, and performed in the Fourier domain for each map, i.e., $\hat {\tilde S}^\mu \approx \tilde {\mathcal{S}}^\mu/\tilde K$. To avoid erroneously emphasizing small scale structures that are inaccessible to EHT and future ground based instruments, we impose a floor on $K$ of $\bar{\tilde {K}}_{10 \ G \lambda}=0.1$ , effectively limiting the deblurring to baselines less than $10\,{\rm G}\lambda$ similar to what is done in practice \citepalias{PaperIII}.

We compare the effective resolutions to two sets of key angular scales.  The first is the resolution of ground-based interferometers at $1.3\,\mm$, roughly $15\,\mu{\rm as}$ with modest super-resolution \citepalias{M87PaperIV,PaperIII}.  Even next-generation ground-based arrays will not be able to significantly improve on this without going to shorter wavelengths (where scattering is less significant).  The second are the major and minor axis sizes of the diffractive scattering kernel, \autoref{eq:scatt_fwhm}, which mark the scale at which diffractive scattering begins to eliminate substructure structure.

\subsection{Kolmogorov Gaussians}
\label{Gaussian Envelopes}

\begin{figure}
    \centering
    \includegraphics[width = \columnwidth]{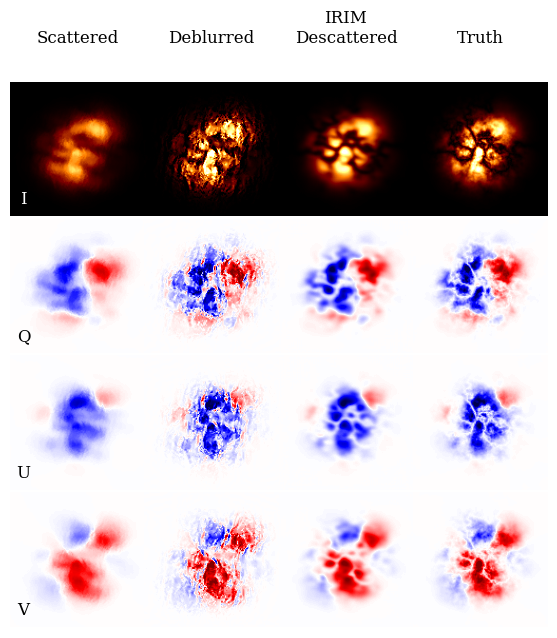}
    \caption{
    Scattered, deblurred, and IRIM descattered images of a representative example test Kolmogorov Gaussian in comparison to the truth for each Stokes parameter.  Deblurring is done using the diffractive scattering kernel.  The shown Kolmogorov Gaussian is an independent realization of the random brightness fluctuations from any in the training or validation sets. Each panel shows a field of view of $200\,\mu{\rm as}$.}
    \label{ehtim-blob}
\end{figure}

\begin{figure}
    \centering
    \includegraphics[width = \columnwidth]{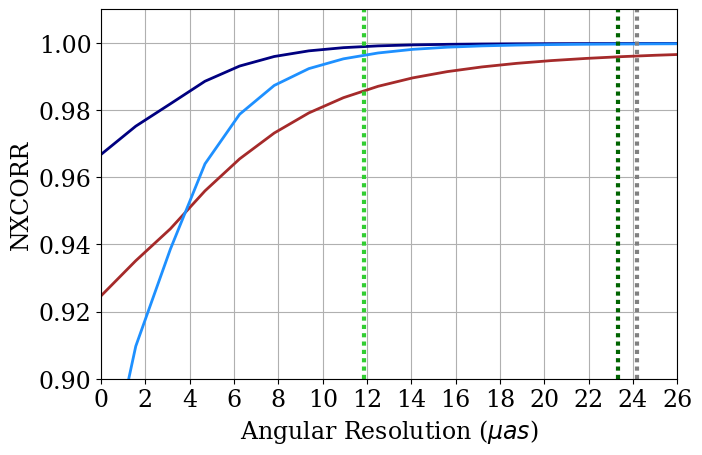} 

    \caption{
    Stokes-parameter averaged NXCORR for a test Kolmogorov Gaussian between the IRIM descattered image (dark blue) in \autoref{ehtim-blob} and the truth after blurring with a Gaussian kernel as a function of the FWHM of the blurring kernel (i.e., angular resolution).  For comparison, similar NXCORRs are shown are the deblurred (light blue) and scattered (red) images.  For reference, the angular scales associated with the diffractive scattering kernel (${\rm FWHM}_{\rm maj}$ and ${\rm FWHM}_{\rm min}$ in dark and light green, respectively), and the instrumental effective imaging resolution of EHT \citepalias[][grey]{M87PaperIV,PaperIII} are indicated by vertical dotted lines.  An NXCORR of unity implies perfect mitigation of the scattering.
    }
    \label{blob-comp}
\end{figure}

We first analyze the performance of the IRIM model on another set of Kolmogorov Gaussians, constructed as described in \autoref{sec:training_set}, though with different realizations of the random fields from those used to train the model.  The four Stokes maps are shown; a representative example is shown in \autoref{ehtim-blob}.  For comparison, the truth, scattered, and deblurred maps of each Stokes parameter are also shown.

The suppression of small-scale structures due to diffractive scattering is immediately evident (as was the case in \autoref{fig:scattering-label}).  Deblurring concentrates the flux, though at the expense of erroneously amplifying refractive substructures.  In contrast, the IRIM model collects the flux diffused by diffractive scattering while simultaneously suppressing the refractive substructures.  The result is an image that is noticeably more similar to the truth.  Nevertheless, as anticipated physically, diffractive scattering limits the capacity of the IRIM model to capture small-scale structures.

The Stokes-parameter-averaged NXCORR between truth and the IRIM model estimate are shown in \autoref{blob-comp}, together with the other comparisons.  At all angular scales, the IRIM model outperforms deblurring and the scattered image, quantitatively confirming the qualitatively impressions from inspecting individual cases.  
At all FWHM explored, the NXCORR of the IRIM model reconsturctions surpass the fiducial threshold of $\rho=0.95$.
Importantly, this extends well below the two sets of key angular scales for ground-based studies of \sgra: the resolution of EHT and future instruments and the scale of the diffractive scattering kernel at $1.3\,\mm$.  Hence, the IRIM model is capable of mitigating scattering to the point that it ceases to be a significant impediment to image reconstruction.

\begin{figure*}
    \centering
    \includegraphics[width=0.39\textwidth]{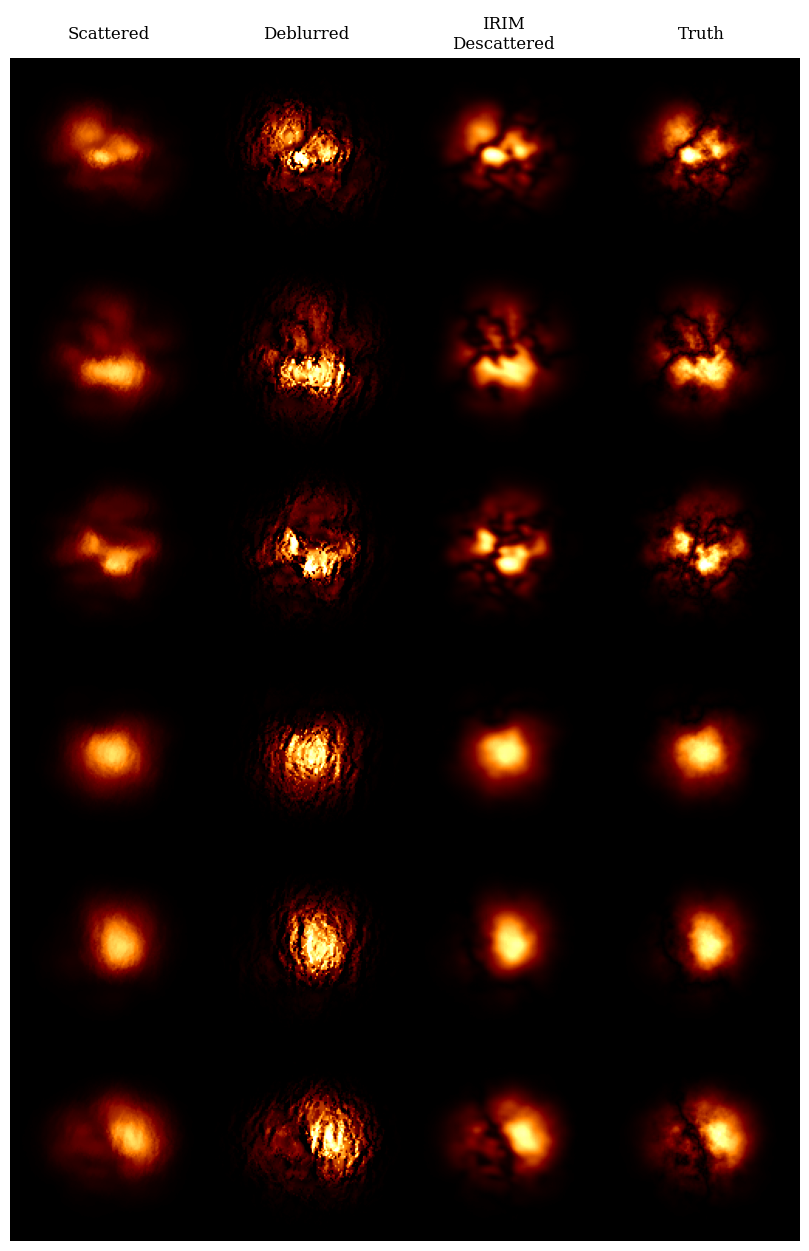}
    \hspace{0.03125in}
    \unskip\ \vrule\
    \hspace{0.03125in}
    \includegraphics[width=0.39\textwidth]{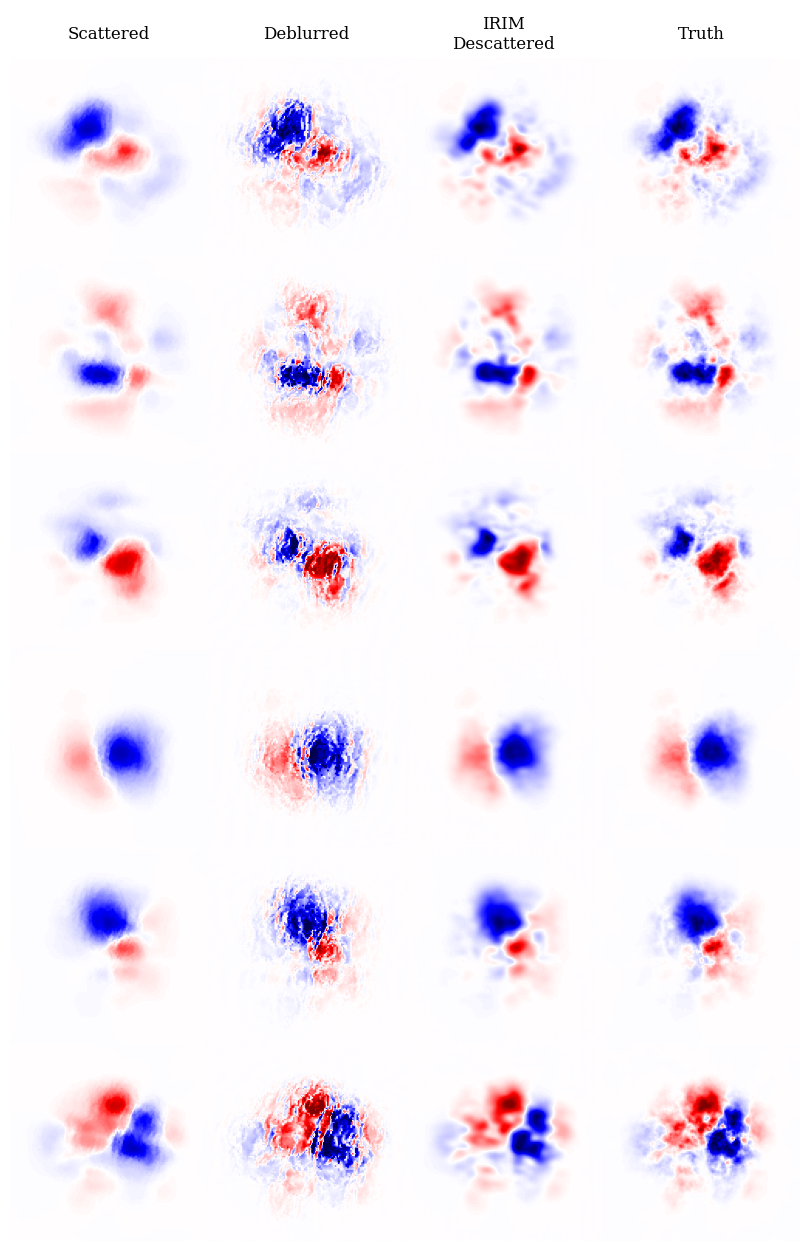}
    \caption{An additional set of 6 distinct Scattered Kolmogorov Gaussians are shown along with their  deblurred, and IRIM descattered reconstructions. The left are positive definite realizations of the Kolmogorov Gaussians used to create the training set Stokes I parameters, on the right are example non-positive-definite realizations, i.e., Stokes Q. These have the field of view of $200 \, \mu{\rm as}$.}
    \label{Test Images}
\end{figure*}

The success of the IRIM model extends to all additional realizations of Kolmogorov Gaussians that we explored, a subset of which are shown in \autoref{Test Images}. That is, the IRIM descattering outperforms deblurring and substantially mitigates the impact of scattering across all accessible angular scales.

\subsection{GRMHD Simulations}

\begin{figure}
    \centering
    \includegraphics[width = 0.999\columnwidth]{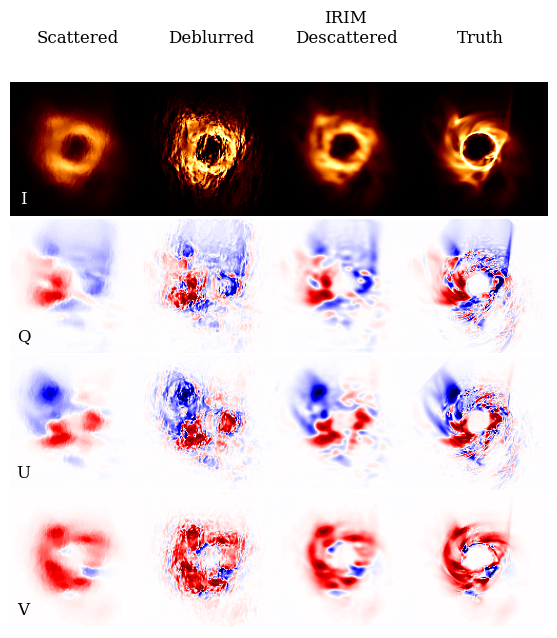}
    \caption{Scattered, deblurred, and Irim descattered images of an example of test GRMHD Images of Sgr A* from \autoref{tab:GRMHDSynthData}. The polarizations are scaled by a factor to the same scale as the training data, allowing for more efficient scattering mitigation. These are the same images as \autoref{fig:scattering-label}, with a field of view of $200 \, \mu{\rm as}$ for each panel.}
    \label{ehtim-sim}
\end{figure}

\begin{figure}
    \centering
\includegraphics[width = \columnwidth]{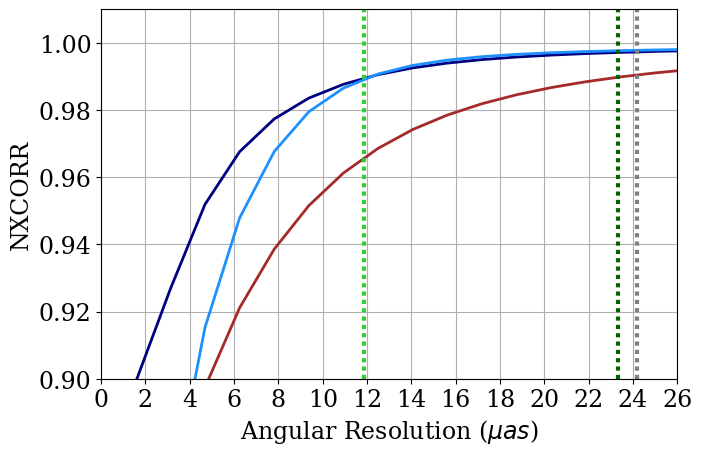}
\caption{Stokes-parameter averaged NXCORR for a test GRMHD simulation between the IRIM descattered image (dark blue), deblurred, and descattered images after blurring with a Gaussian kernel as a function of the FWHM blurring kernel\autoref{ehtim-sim}. 
Gaussian between the IRIM descattered image
(dark blue) in Figure 3 and the truth after blurring with
a Gaussian kernel as a function of the FWHM of the angular resolution. The deblurred (light blue) and scattered (red) images are compared to the IRIM descattered. The angular scales (dashed lines) associated with the diffractive scattering kernel are $\text{FWHM}_{\text{maj}}$ (dark green), $\text{FWHM}_{\text{min}}$ (light green). The instrumental effective resolution of EHT (\citetalias{M87PaperIV},\citetalias{PaperIII}; grey). }
    \label{sim-comp}
\end{figure}

We repeat the appraisal of the performance of IRIM model with a set of images from GRMHD simulations of black hole accretion flows, taken from the simulation suite in \citetalias{M87PaperV}. The polarizations are scaled to the same scale as the Kolmogorov Gaussians, $I \in [0,1]$ and $Q, \ U, \ V \in [-1,1]$. This allows the model to mitigate scattering more effectively. 
The generation of these images and the physics of the underlying simulations are described in detail in \citetalias{M87PaperV,PaperV} and \citet{GRMHD_comparison:2019}, and only very briefly summarized here.  They self-consistently solve the MHD equations on a Kerr background, yielding the time-evolution of the density, velocity, and magnetic field of a geometrically thick accretion flow.  They are subsequently post-processed to generate the distribution of hot electrons, using a plasma-$\beta$ dependent prescription for the electron-ion temperature ratio, and then images of the subsequent synchrotron emission.  

We use a set of independent snapshot images taken from the validation simulation sets from \citetalias{PaperIV}, corresponding to simulation data sets 90-99 in Table 7 therein.

While these images share the typical scale of the training set images (by construction of the latter), they qualitatively differ in number of ways.  First, they have a clear ring-like morphology associated with the shadow produced by strong gravitational lensing about the black hole and presence of an event horizon.  Second, the fluctuations are self-consistently generated and associated with the turbulence and propagating features in the accretion flow.  As a consequence, the fluctuations have a physically significant power spectrum, and more importantly, clearly exhibit correlations not present in the training set images.  Third, the emission mechanism imposes significant correlations among the Stokes maps associated with the underlying magnetic field geometry and bulk motions within the accretion flow.  

\autoref{sim-comp} presents the truth and IRIM-descattered Stokes maps for a representative GRMHD simulation (MAD flow type, $R_{\rm high}=160$, $i=30^\circ$, and $a_*=-0.5$; we direct the reader to \citetalias{PaperV} for definitions of these parameters).
Again, for comparison we also show the Stokes maps for the scattered and deblurred image.  As in \autoref{Gaussian Envelopes}, deblurring produces artificial features associated with the refractive scattering.

Neither of these corruptions are present in the IRIM model descattered images, which accurately recover many of the spiral structures in the truth image.

The Stokes-parameter-averaged NXCORRs, shown in \autoref{ehtim-sim}, reflect the greater fidelity of the IRIM model estimates of the intrinsic image.  
The additional refractive substructures in the deblurred images are heavily penalized at small angular scales.
Therefore, as with the Kolmogorov Gaussians, the IRIM model out performs deblurring at all angular scales, being larger than the fiducial threshold of $\rho=0.95$ for all scales larger than $4\,\mu{\rm as}$.  

\begin{deluxetable}{ccCcc}
\tabletypesize{\normalsize}
\tablewidth{0pt}
\tablecaption{GRMHD synthetic data set parameters \label{tab:GRMHDSynthData}}

\tablehead{
\colhead{\begin{minipage}{3cm}
\centering
\citetalias{PaperIV}\\
Index\tablenotemark{a}
\end{minipage}}
& \colhead{
\begin{minipage}{1.5cm}
\centering
Accretion\\
State
\end{minipage}} 
& \colhead{$a_*$} 
& \colhead{
\begin{minipage}{1cm}
\centering
$i$\\
(deg)
\end{minipage}}
& \colhead{$R_{\text{high}}$}\\[-0.5cm]
}
\startdata
092 & MAD  & -0.5\hphantom{0}              & 30  & 160 \\
\hline
090 & MAD  & \hphantom{-}0.0\hphantom{0}   & 150 & 160 \\
091 & MAD  & \hphantom{-}0.5\hphantom{0}   & 70  & 160 \\
093 & MAD  & \hphantom{-}0.94              & 30  & 10  \\
095 & SANE & -0.94                         & 70  & 10  \\
096 & SANE & \hphantom{-}0.5\hphantom{0}   & 110 & 40  \\
098 & SANE & \hphantom{-}0.0\hphantom{0}   & 150 & 40 \\
\enddata
\tablenotetext{a}{Index of simulation in Table 7 of \citetalias{PaperIV}.}
\tablecomments{Simulation parameters for each of the GRMHD-based synthetic data sets from which the test images in \autoref{ehtim-sim} (above the line) and \autoref{GRMHD} (in order below the line).  These were used in the validation of the EHT \sgra mass measurement in \citetalias{PaperIV}, where more information may be found.}
\end{deluxetable}

\begin{figure*}
    \begin{center}
    \includegraphics[width=0.4\textwidth]{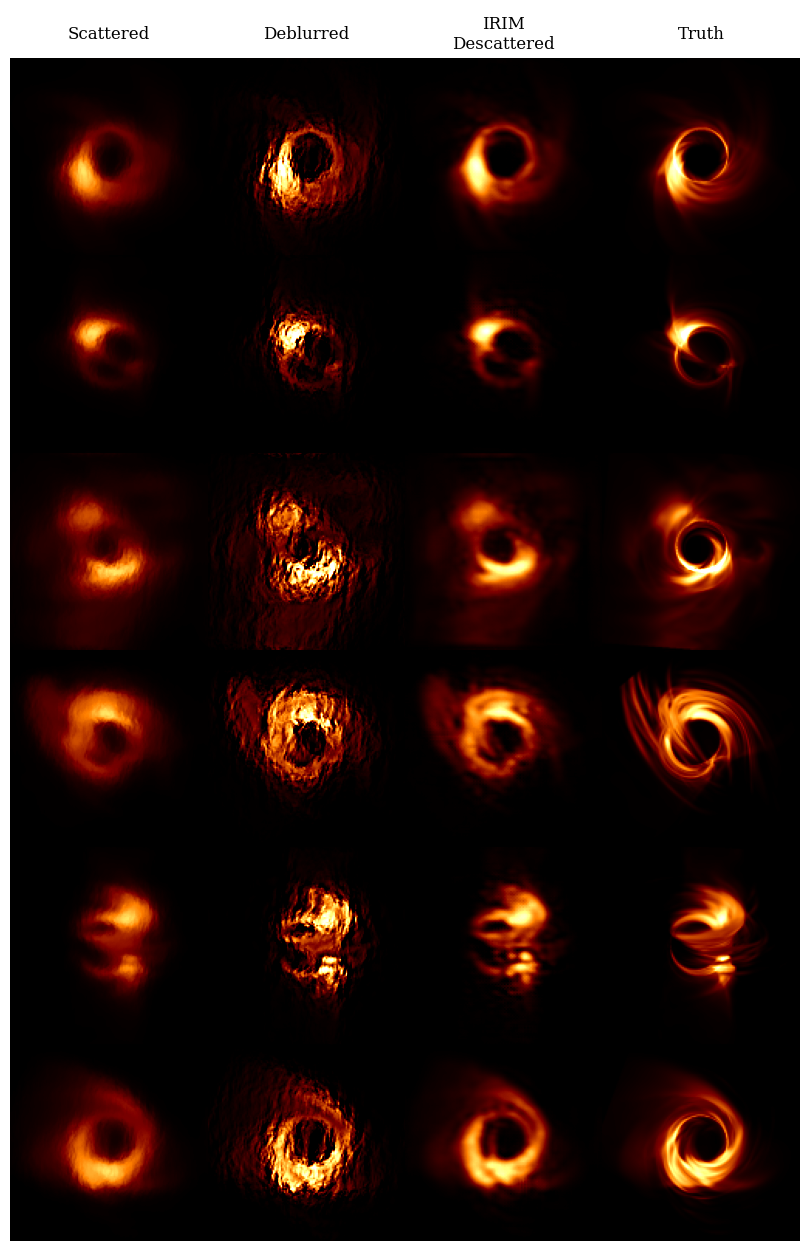}
    \hspace{0.03125in}
    \unskip\ \vrule\
    \hspace{0.03125in}
    \includegraphics[width=0.4\textwidth]{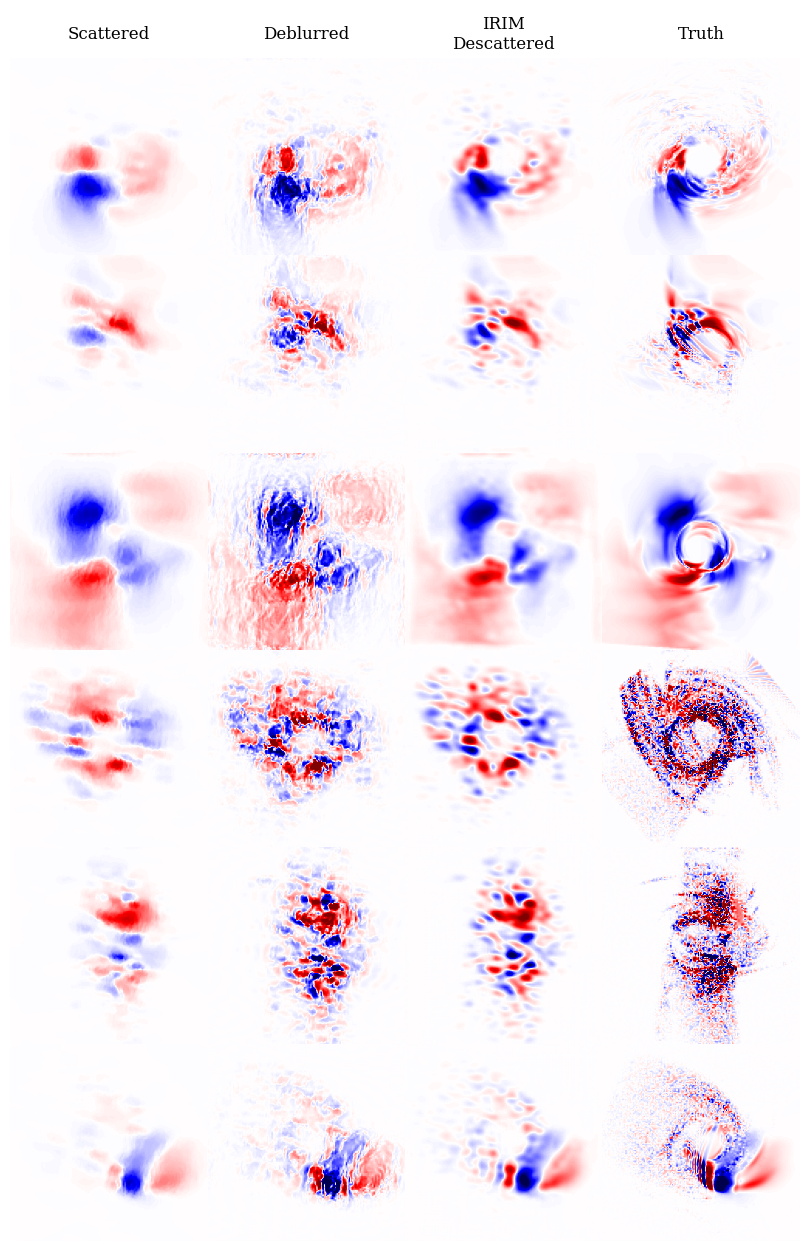}
    \includegraphics[width=0.4\textwidth]{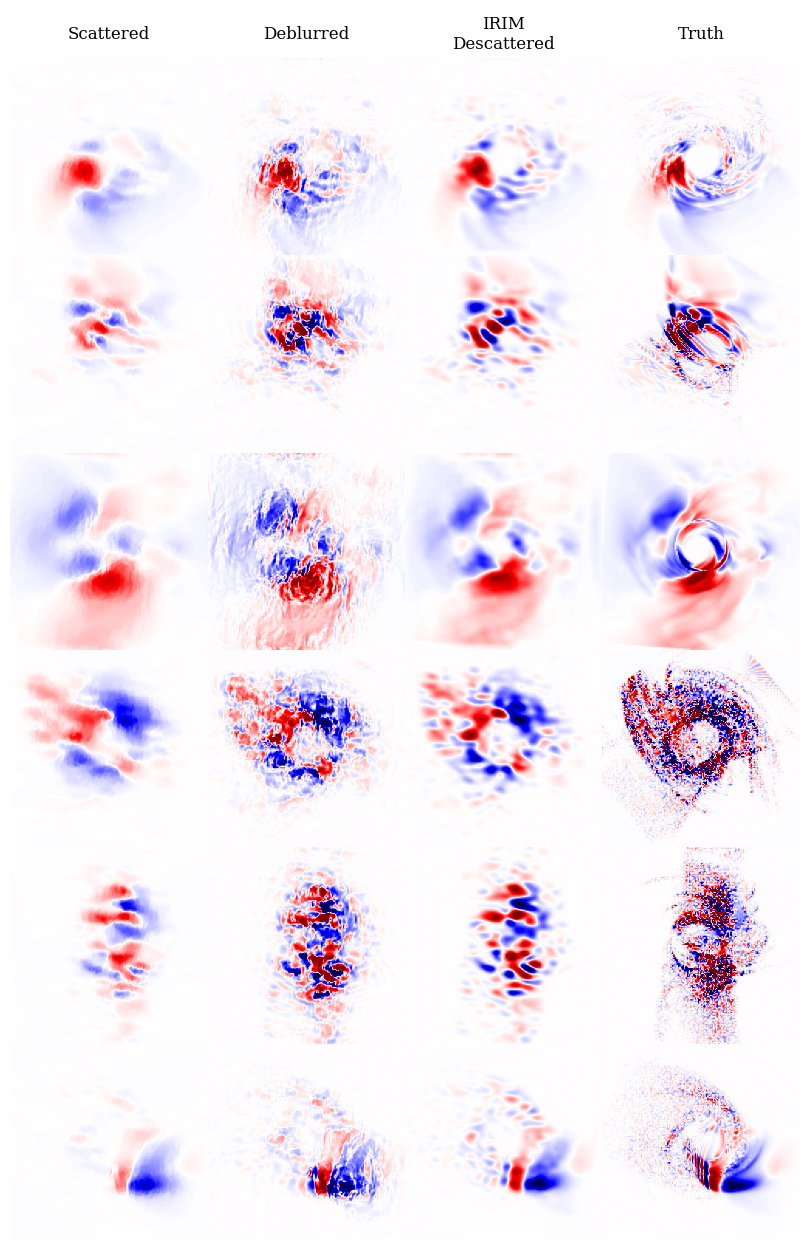}
    \hspace{0.03125in}
    \unskip\ \vrule\
    \hspace{0.03125in}
    \includegraphics[width=0.4\textwidth]{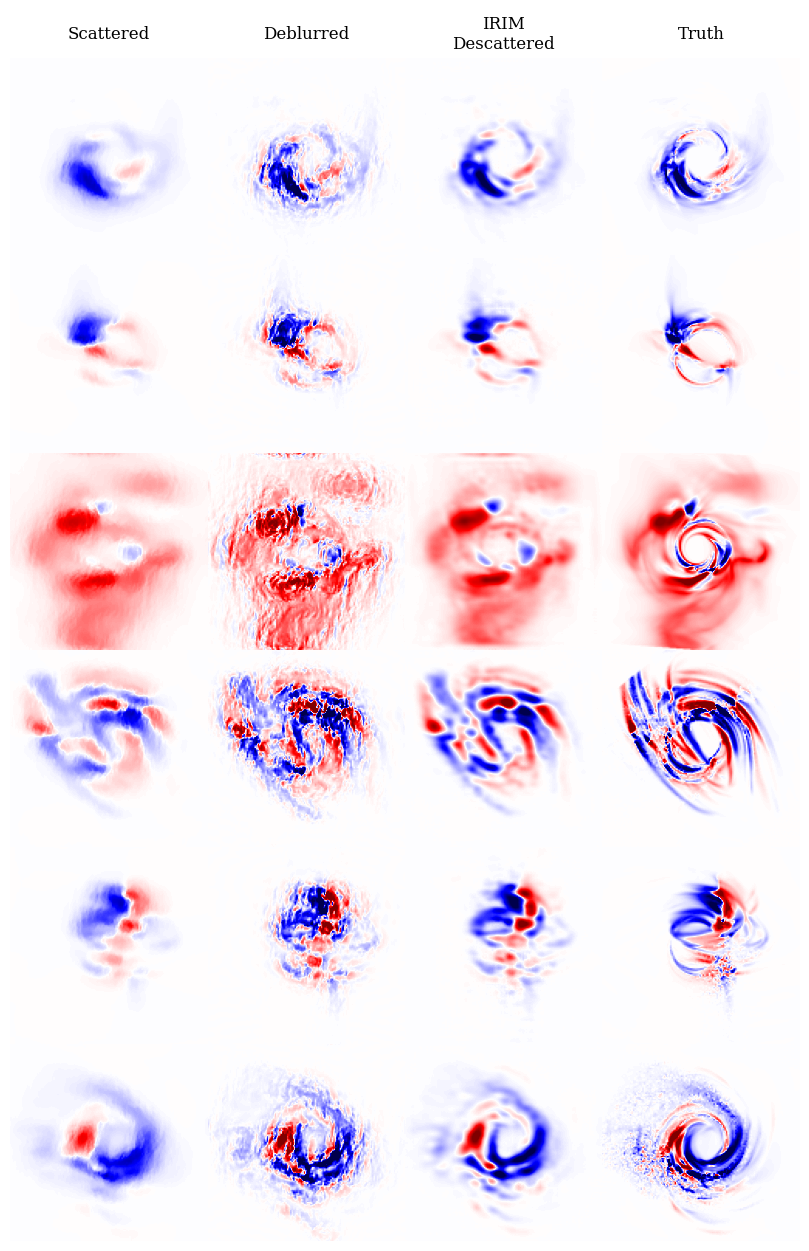}
    \end{center}
    \caption{An additional set of scattered, deblurred, and Irim descattered images of an example of test GRMHD Images of Sgr A* from \autoref{tab:GRMHDSynthData}. Similar to \autoref{ehtim-sim} the polarizations are scaled to the same scales as the Kolmogorov Gaussians to improve the models ability to mitigate scattering. The field of view for the polarization panels are $200 \, \mu{\rm as}$. }
    \label{GRMHD}
\end{figure*}

Similar results are found for all of the GRMHD simulations that we have inspected, a subset of which are listed in \autoref{tab:GRMHDSynthData} and shown in \autoref{GRMHD}.  In all instances, the IRIM model estimate outperforms the deblurred and the scattered image substantially for all Stokes maps.  That is, it appears that it is generically possible to mitigate scattering to well below the resolution of EHT and future ground-based interferometers with IRIM.  Importantly, this appears to be possible without introducing erroneous artifacts associated with the refractive scattering.

\section{Conclusions} 
\label{Conclusion}

Interstellar scattering may be described as a convolution of a noisy kernel and the intrinsic brightness maps.  For relevant astrophysical parameters, the scattering kernel is nonbirefringent, i.e., it is independent of Stoke parameter.  Thus, the observing multiple stokes maps presents multiple images convolved with the same realization of the scattering screen.  By leveraging this fact, we have demonstrated the ability to effectively mitigate scattering at $1.3\,\mm$, despite deconvolution being formally poorly defined.

The machine learning IRIM formalism, developed to denoise images, closely matches the nature of the scattering problem.  An IRIM descattering model trained on a purely phenomenological set of mock images effectively mitigates scattering in physically motivated images from GRMHD simulations.  Importantly, these simulations differ from the training set in many aspects, including correlations among Stokes maps, correlations between stochastic features within Stokes maps, and the clear presence of a shadow resulting in a ring-like morphology.  Nevertheless, the IRIM descattering significantly outperforms deblurring with the diffractive scattering kernel, currently the most commonly used mitigation scheme (\citetalias{PaperIII,PaperIV};\citealt{Broderick_2022}).

The success of the IRIM descattering model provides direct evidence that sufficient information exists within the scattered images to identify and remove the impacts of both diffractive and refractive scattering on resolutions well beyond those practically accessible in the near future.  That is, the well-posedness of the deconvolution problem is not a substantial impediment to practical scattering mitigation.  This success motivates future development of scattering mitigation schemes to be used as a part of the image reconstruction pipelines for EHT, ngEHT and other future mm-wavelength interferometers.  Efforts to do so must contend with a number of additional complications, including instrument resolution (i.e., $(u,v)$-coverage), additional sources of measurement error (thermal noise, systematic uncertainties), and the particulars of individual imaging methods.  For these reasons, we defer such development to future work.

At the same time, any practical pipeline could benefit from a number of additional modifications.  First and foremost, as $(u,v)$-coverage improves, it will become possible to generate movies of \sgra, presenting many independent realizations of the emission region for an essentially fixed scattering screen.  Second, simultaneous observations at multiple wavelengths (e.g., $3\,\mm$, $1.3\,\mm$, and $0.86\,\ mm$) present the opportunity to leverage the differing spectral dependencies of the emission and scattering to separate the two.  Third, we have opted to train on a particularly pessimistic set of images.  The Kolmogorov Gaussians do not have the ringlike structure prominent in EHT observations, do not have strong correlations among the Stokes maps anticipated by radiative transfer calculations, and do not have spatial correlations among the intrinsic image structures.  In addition, we adopted a Kolmogorov spectrum for the fluctuations, similar to that for the phase fluctuations that define the scattering screen, and minimizing the disparity between refractively generated substructures and those in the intrinsic image.  Incorporating any of these properties into the training would improve the capacity to isolate the scattering screen and mitigate its effects.

\section{Acknowledgements}

We thank Ben Prather, Abhishek Joshi, Vedant Dhruv, C.K. Chan, and Charles Gammie for the synthetic images used here, generated under NSF grant AST 20-34306.  This research used resources of the Oak Ridge Leadership Computing Facility at the Oak Ridge National Laboratory, which is supported by the Office of Science of the U.S.\ Department of Energy under Contract No. DE-AC05-00OR22725. This research used resources of the Argonne Leadership Computing Facility, which is a DOE Office of Science User Facility supported under Contract DE-AC02-06CH11357. This research was done using services provided by the OSG Consortium, which is supported by the National Science Foundation awards \#2030508 and \#1836650. This research is part of the Delta research computing project, which is supported by the National Science Foundation (award OCI 2005572), and the State of Illinois. Delta is a joint effort of the University of Illinois at Urbana-Champaign and its National Center for Supercomputing Applications.
We would like to thank Dr.\ Robert Mann for suggesting this collaboration.
This work was supported in part by Perimeter Institute for Theoretical Physics.  Research at Perimeter Institute is supported by the Government of Canada through the Department of Innovation, Science and Economic Development Canada and by the Province of Ontario through the Ministry of Economic Development, Job Creation and Trade.
A.E.B. receives additional financial support from the Natural Sciences and Engineering Research Council of Canada through a Discovery Grant.

\bibliography{main,EHTCpapers}{}
\bibliographystyle{aasjournal}

\end{document}